\documentclass[12pt,a4paper]{article}

\usepackage{ifthen} 
\newboolean{pdflatex}
\setboolean{pdflatex}{true} 

\newboolean{articletitles}
\setboolean{articletitles}{true} 

\newboolean{uprightparticles}
\setboolean{uprightparticles}{false} 

\newboolean{inbibliography}
\setboolean{inbibliography}{false} 

\usepackage[top=1in, bottom=1.25in, left=1in, right=1in]{geometry}

\usepackage{microtype}
\usepackage{lineno}  
\usepackage{xspace} 
\usepackage{caption} 

\usepackage{graphicx}  
\usepackage{color}
\usepackage{colortbl}
\graphicspath{{./figs/}} 

\usepackage{amsmath} 
\usepackage{amssymb}
\usepackage{amsfonts}
\usepackage{upgreek} 

\newcommand*\patchAmsMathEnvironmentForLineno[1]{%
\expandafter\let\csname old#1\expandafter\endcsname\csname #1\endcsname
\expandafter\let\csname oldend#1\expandafter\endcsname\csname
end#1\endcsname
 \renewenvironment{#1}%
   {\linenomath\csname old#1\endcsname}%
   {\csname oldend#1\endcsname\endlinenomath}%
}
\newcommand*\patchBothAmsMathEnvironmentsForLineno[1]{%
  \patchAmsMathEnvironmentForLineno{#1}%
  \patchAmsMathEnvironmentForLineno{#1*}%
}
\AtBeginDocument{%
\patchBothAmsMathEnvironmentsForLineno{equation}%
\patchBothAmsMathEnvironmentsForLineno{align}%
\patchBothAmsMathEnvironmentsForLineno{flalign}%
\patchBothAmsMathEnvironmentsForLineno{alignat}%
\patchBothAmsMathEnvironmentsForLineno{gather}%
\patchBothAmsMathEnvironmentsForLineno{multline}%
\patchBothAmsMathEnvironmentsForLineno{eqnarray}%
}

\usepackage{hyperref}    
\usepackage[all]{hypcap} 

\usepackage{xspace} 
\usepackage{upgreek}

\def\lhcb {\mbox{LHCb}\xspace}

\def\MagUp {\mbox{\em Mag\kern -0.05em Up}\xspace}

\ifthenelse{\boolean{uprightparticles}}%
{

 \def\PDelta      {\ensuremath{\Delta}\xspace}                 
 \def\PXi      {\ensuremath{\Xi}\xspace}                 
 \def\PLambda      {\ensuremath{\Lambda}\xspace}                 
 \def\PSigma      {\ensuremath{\Sigma}\xspace}                 
 \def\POmega      {\ensuremath{\Omega}\xspace}                 
 \def\PUpsilon      {\ensuremath{\Upsilon}\xspace}                 
 
 \def\PB      {\ensuremath{\mathrm{B}}\xspace}                 
                  
 \def\PD      {\ensuremath{\mathrm{D}}\xspace}

 \def\PK      {\ensuremath{\mathrm{K}}\xspace}

 \def\Pb      {\ensuremath{\mathrm{b}}\xspace}                 
 \def\Pc      {\ensuremath{\mathrm{c}}\xspace}

 \def\Pi      {\ensuremath{\mathrm{i}}\xspace}

 \def\Ps      {\ensuremath{\mathrm{s}}\xspace}

}
{

 \mathchardef\PDelta="7101
 \mathchardef\PXi="7104
 \mathchardef\PLambda="7103
 \mathchardef\PSigma="7106
 \mathchardef\POmega="710A
 \mathchardef\PUpsilon="7107
                  
 \def\PB      {\ensuremath{B}\xspace}                 
                  
 \def\PD      {\ensuremath{D}\xspace}

 \def\PK      {\ensuremath{K}\xspace}

 \def\Pb      {\ensuremath{b}\xspace}                 
 \def\Pc      {\ensuremath{c}\xspace}

 \def\Pi      {\ensuremath{i}\xspace}

 \def\Ps      {\ensuremath{s}\xspace}

}

\makeatletter
\ifcase \@ptsize \relax
  \newcommand{\miniscule}{\@setfontsize\miniscule{4}{5}}
\or
  \newcommand{\miniscule}{\@setfontsize\miniscule{5}{6}}
\or
  \newcommand{\miniscule}{\@setfontsize\miniscule{5}{6}}
\fi
\makeatother

\DeclareRobustCommand{\optbar}[1]{\shortstack{{\miniscule (\rule[.5ex]{1.25em}{.18mm})}
  \\ [-.7ex] $#1$}}

\def\squark    {{\ensuremath{\Ps}}\xspace}

\def\cquark    {{\ensuremath{\Pc}}\xspace}

\def\bquark    {{\ensuremath{\Pb}}\xspace}

\def\kaon    {{\ensuremath{\PK}}\xspace}
  \def\Kbar    {{\kern 0.2em\overline{\kern -0.2em \PK}{}}\xspace}

\def\KorKbar    {\kern 0.18em\optbar{\kern -0.18em K}{}\xspace}

\def\KS      {{\ensuremath{\kaon^0_{\mathrm{ \scriptscriptstyle S}}}}\xspace}

  \def\Dbar    {{\kern 0.2em\overline{\kern -0.2em \PD}{}}\xspace}
\def\D       {{\ensuremath{\PD}}\xspace}

\def\DorDbar    {\kern 0.18em\optbar{\kern -0.18em D}{}\xspace}

\def\Dp      {{\ensuremath{\D^+}}\xspace}

\def\Ds      {{\ensuremath{\D^+_\squark}}\xspace}
\def\Dsp     {{\ensuremath{\D^+_\squark}}\xspace}

\def\Bbar    {{\ensuremath{\kern 0.18em\overline{\kern -0.18em \PB}{}}}\xspace}

\def\BorBbar    {\kern 0.18em\optbar{\kern -0.18em B}{}\xspace}

  \def\Y#1S{\ensuremath{\PUpsilon{(#1S)}}\xspace}

\def\Lbar        {{\ensuremath{\kern 0.1em\overline{\kern -0.1em\PLambda}}}\xspace}
\def\LorLbar    {\kern 0.18em\optbar{\kern -0.18em \PLambda}{}\xspace}

\def\to                 {\ensuremath{\rightarrow}\xspace}

\def\CP                {{\ensuremath{C\!P}}\xspace}

\def\AT#1     {\ensuremath{A_{\mathrm{T}}^{#1}}\xspace}           

\def\C#1      {\ensuremath{\mathcal{C}_{#1}}\xspace}                       
\def\Cp#1     {\ensuremath{\mathcal{C}_{#1}^{'}}\xspace}                    
\def\Ceff#1   {\ensuremath{\mathcal{C}_{#1}^{\mathrm{(eff)}}}\xspace}        
\def\Cpeff#1  {\ensuremath{\mathcal{C}_{#1}^{'\mathrm{(eff)}}}\xspace}       
\def\Ope#1    {\ensuremath{\mathcal{O}_{#1}}\xspace}                       
\def\Opep#1   {\ensuremath{\mathcal{O}_{#1}^{'}}\xspace}                    

\newcommand{\tev}{\ifthenelse{\boolean{inbibliography}}{\ensuremath{~T\kern -0.05em eV}\xspace}{\ensuremath{\mathrm{\,Te\kern -0.1em V}}}\xspace}
\newcommand{\gev}{\ensuremath{\mathrm{\,Ge\kern -0.1em V}}\xspace}
\newcommand{\mev}{\ensuremath{\mathrm{\,Me\kern -0.1em V}}\xspace}
\newcommand{\kev}{\ensuremath{\mathrm{\,ke\kern -0.1em V}}\xspace}
\newcommand{\ev}{\ensuremath{\mathrm{\,e\kern -0.1em V}}\xspace}
\newcommand{\gevc}{\ensuremath{{\mathrm{\,Ge\kern -0.1em V\!/}c}}\xspace}
\newcommand{\mevc}{\ensuremath{{\mathrm{\,Me\kern -0.1em V\!/}c}}\xspace}
\newcommand{\gevcc}{\ensuremath{{\mathrm{\,Ge\kern -0.1em V\!/}c^2}}\xspace}
\newcommand{\gevgevcccc}{\ensuremath{{\mathrm{\,Ge\kern -0.1em V^2\!/}c^4}}\xspace}
\newcommand{\mevcc}{\ensuremath{{\mathrm{\,Me\kern -0.1em V\!/}c^2}}\xspace}

\def\mum  {\ensuremath{{\,\upmu\mathrm{m}}}\xspace}

\def\gsim{{~\raise.15em\hbox{$>$}\kern-.85em
          \lower.35em\hbox{$\sim$}~}\xspace}
\def\lsim{{~\raise.15em\hbox{$<$}\kern-.85em
          \lower.35em\hbox{$\sim$}~}\xspace}

\def\ptot       {\mbox{$p$}\xspace}
\def\pt         {\mbox{$p_{\mathrm{ T}}$}\xspace}

\def\evtgen     {\mbox{\textsc{EvtGen}}\xspace}

\def\geant      {\mbox{\textsc{Geant4}}\xspace}

\def\photos     {\mbox{\textsc{Photos}}\xspace}

\def\pythia     {\mbox{\textsc{Pythia}}\xspace}

\def\tell1  {TELL1\xspace}
\def\ukl1   {UKL1\xspace}

\newcommand{\ie}{\mbox{\itshape i.e.}\xspace}

\newcommand{\epp}{\ensuremath{\eta\to \pi^+\pi^-}\xspace}
\newcommand{\eppp}{\ensuremath{\eta^\prime\to \pi^+\pi^-}\xspace}
\newcommand{\epppp}{\ensuremath{\eta^\prime(958)\to \pi^+\pi^-}\xspace}
\newcommand{\etaspp}{\ensuremath{\eta^{({\prime})}\to \pi^+\pi^-}\xspace}
\newcommand{\etas}{\ensuremath{\eta^{({\prime})}}\xspace}
\newcommand{\dppp}{\ensuremath{D^+\to \pi^+\pi^+\pi^-}\xspace}

\newcommand{\dsppp}{\ensuremath{D_s^+\to \pi^+\pi^+\pi^-}\xspace}
\newcommand{\dpsppp}{\ensuremath{D_{(s)}^+\to \pi^+\pi^+\pi^-}\xspace}

\newcommand{\BRof}[1]{\ensuremath{{\cal B}\left(#1\right)}\xspace}
\newcommand{\pp}{\ensuremath{\pi^+\pi^-}\xspace}

\newcommand{\threepi}{\ensuremath{\pi^+ \pi^+ \pi^-}\xspace}

\newcommand{\pippim}{\ensuremath{\pi^+\pi^-}}

\newcommand{\CLs}{\ensuremath{\textrm{CL}_{\textrm{s}}}\xspace}

\usepackage{longtable} 

\usepackage{cite} 
\usepackage{mciteplus}
\begin{document}

\renewcommand{\thefootnote}{\fnsymbol{footnote}}
\setcounter{footnote}{1}

\begin{titlepage}
\pagenumbering{roman}

\vspace*{-1.5cm}
\centerline{\large EUROPEAN ORGANIZATION FOR NUCLEAR RESEARCH (CERN)}
\vspace*{1.5cm}
\noindent
\begin{tabular*}{\linewidth}{lc@{\extracolsep{\fill}}r@{\extracolsep{0pt}}}
\ifthenelse{\boolean{pdflatex}}
{\vspace*{-2.7cm}\mbox{\!\!\!\includegraphics[width=.14\textwidth]{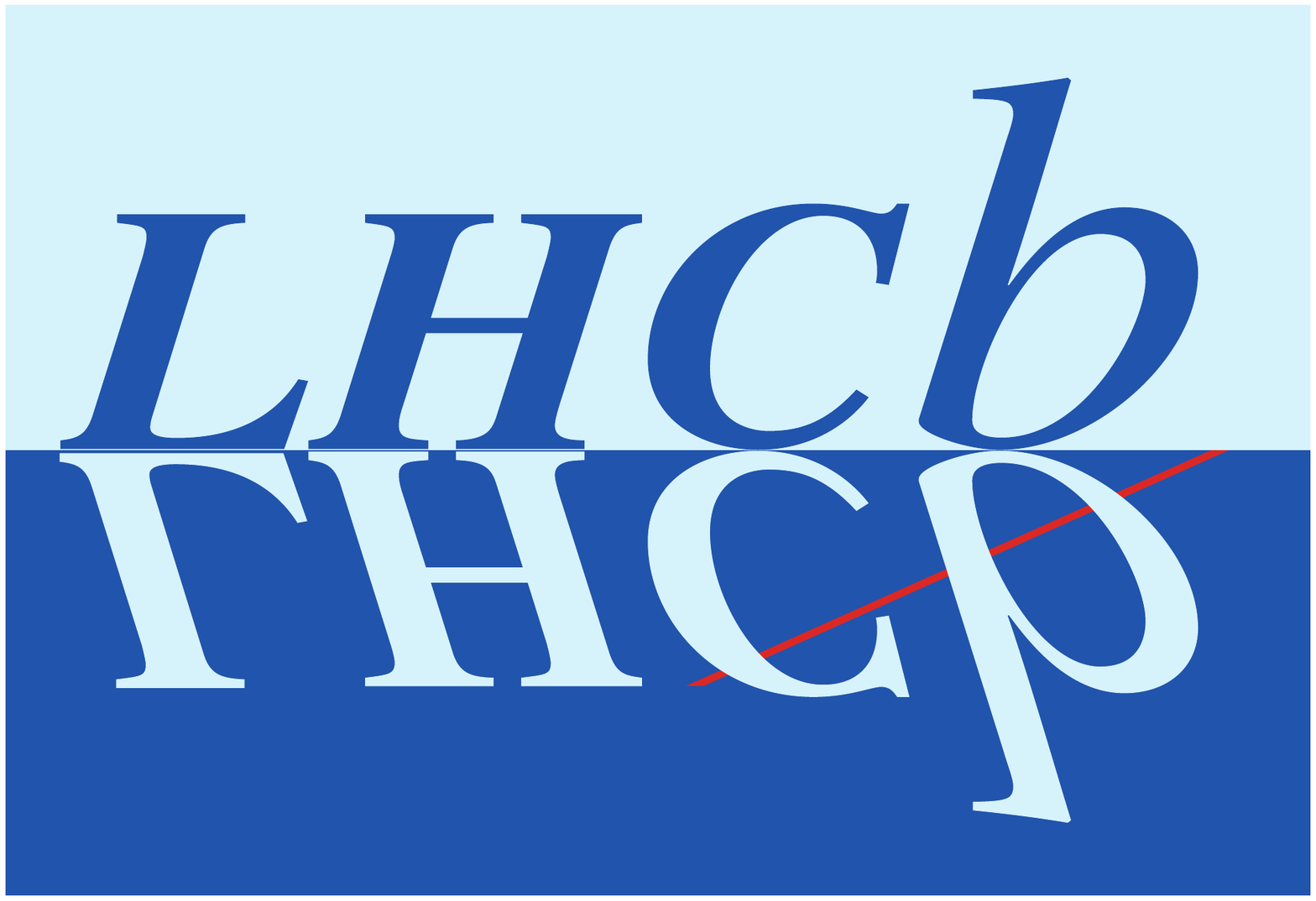}} & &}%
{\vspace*{-1.2cm}\mbox{\!\!\!\includegraphics[width=.12\textwidth]{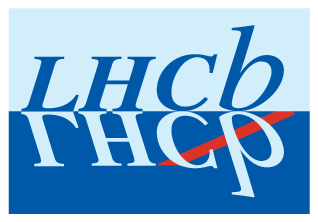}} & &}%
\\
 & & CERN-EP-2016-257 \\  
 & & LHCb-PAPER-2016-046 \\  
 & & October 12, 2016 \\ 
 \\
\end{tabular*}

\vspace*{3.0cm}

{\normalfont\bfseries\boldmath\huge
\begin{center}
  Search for the \CP-violating strong decays $\eta \to \pi^+\pi^-$ and $\eta^\prime(958) \to \pi^+\pi^-$
\end{center}
}

\vspace*{1.0cm}

\begin{center}
The LHCb collaboration\footnote{Authors are listed at the end of this letter.}
\end{center}

\vspace{\fill}

\begin{abstract}
  \noindent
 A search for the \CP-violating strong decays $\eta \to \pi^+\pi^-$ and $\eta^\prime(958) \to \pi^+\pi^-$ has been performed using approximately 
 $2.5 \times 10^{7}$ events of each of the decays $D^+ \to \pi^+\pi^+\pi^-$ 
 and $D_s^+ \to \pi^+\pi^+\pi^-$, recorded by the LHCb experiment. The data set corresponds to an integrated luminosity of 3.0\,fb$^{-1}$ of $pp$ collision data recorded during LHC Run~1 and 0.3\,fb$^{-1}$ recorded in Run~2. 
 No evidence is seen for 
 $D^+_{(s)} \to \pi^+ \eta^{(\prime)}$ with $\eta^{(\prime)} \to \pi^+\pi^-$,
 and upper limits at
 90\% confidence level are set on the branching fractions,
 $\mathcal{B}(\eta \to \pi^+\pi^-) < 1.6 \times 10^{-5}$ and 
 $\mathcal{B}(\eta^\prime \to \pi^+\pi^-) < 1.8 \times 10^{-5}$. The limit for the $\eta$ decay is comparable with the existing 
 one, while that for the $\eta^\prime$ is a factor of three smaller than the previous limit.
  
\end{abstract}

\vspace*{2.0cm}

\begin{center}
  Published in Phys.~Lett.~B 764 (2017) 233-240
\end{center}

\vspace{\fill}

{\footnotesize 
\centerline{\copyright~CERN on behalf of the \lhcb collaboration, licence \href{http://creativecommons.org/licenses/by/4.0/}{CC-BY-4.0}.}}
\vspace*{2mm}

\end{titlepage}


\newpage
\setcounter{page}{2}
\mbox{~}

\cleardoublepage
\renewcommand{\thefootnote}{\arabic{footnote}}
\setcounter{footnote}{0}



\pagestyle{plain} 
\setcounter{page}{1}
\pagenumbering{arabic}

\section{Introduction}
\label{sec:Introduction}

The strength of \CP violation in weak interactions in the quark sector is well 
below what would be required to serve as an explanation 
for the observed imbalance between the amounts of matter and antimatter in the universe. 
The QCD Lagrangian could contain a term, the $\theta$ term~\cite{Cheng:1987gp}, 
that would give rise to \CP violation in strong interactions; 
however, no strong \CP violation has been
observed. 
The experimental upper limit on the neutron 
electric dipole moment ($n$EDM) implies a limit $\theta \lesssim 
10^{-10}$~\cite{Kuckei:2007}. The closeness of the value of $\theta$ to zero 
is seen as a
fine-tuning problem, the so-called ``strong \CP problem''. Solutions to
the strong \CP problem may involve axions~\cite{Peccei:2006as}, extra space-time 
dimensions~\cite{Khlebnikov:1987zg},
massless up quarks~\cite{Nelson:2003tb}, string theory~\cite{Leigh:1993ae} or quantum 
gravity~\cite{Berezhiani:1992pq}.

The decay modes \epp and \epppp would both violate
\CP symmetry.
In the Standard Model (SM) these decays could happen via the 
\CP-violating
weak interaction,
through mediation by a virtual \KS meson, with expected branching fractions
$\BRof{\epp} < 2 \times 10^{-27}$ and 
$\BRof{\eppp} < 4 \times 10^{-29}$~\cite{Jarlskog:1995ww}.
Based on the limit from the $n$EDM measurements, strong
decays mediated by the $\theta$ term would have branching fractions below 
about 
$3 \times 10^{-17}$~\cite{Jarlskog:1995ww}. Any observation of larger 
branching fractions  
would indicate a new source of \CP violation in the strong interaction, which could 
help to solve
the problem of the origin of the matter-antimatter asymmetry.
The current limit for the \epp decay mode, 
$\BRof{\epp}< 1.3 \times 10^{-5}$ at 
90\% confidence level (CL), comes from the
KLOE experiment~\cite{Ambrosino:2004ww}, which 
looked for \epp in the decay $\phi(1020)\to \eta \gamma$.  
The limit for $\eta^\prime$, $\BRof{\eppp}< 5.5 \times 10^{-5}$ at 90\% CL, 
is from the BESIII experiment~\cite{Ablikim:2011vg}, based on searches for
$\eta^\prime \to \pi^+\pi^-$ in radiative $J/\psi \to \eta^\prime \gamma$ decays. 
In the study presented here, a new method is introduced to search for the
decays $\eta \to \pp$ and $\eta^\prime \to \pp$, exploiting the large sample of charm mesons collected by \lhcb.  

\section{Detector and simulation}
\label{sec:Detector}
The \lhcb detector~\cite{Alves:2008zz,LHCb-DP-2014-002} is a single-arm forward
spectrometer covering the \mbox{pseudorapidity} range $2<\eta <5$,
designed for the study of particles containing \bquark or \cquark
quarks. The detector includes a high-precision tracking system
consisting of a silicon-strip vertex detector surrounding the $pp$
interaction region, a large-area silicon-strip detector located
upstream of a dipole magnet with a bending power of about
$4{\mathrm{\,Tm}}$, and three stations of silicon-strip detectors and straw
drift tubes placed downstream of the magnet.
The tracking system provides a measurement of momentum, \ptot, of charged particles with
a relative uncertainty that varies from 0.5\% at low momentum to 1.0\% at 200\gevc.
The minimum distance of a track to a primary $pp$ interaction vertex (PV), the impact parameter, is measured with a resolution of $(15+29/\pt)\mum$,
where \pt is the component of the momentum transverse to the beam, in\,\gevc.
Different types of charged hadrons are distinguished using information
from two ring-imaging Cherenkov detectors. 
Photons, electrons and hadrons are identified by a calorimeter system consisting of
scintillating-pad and preshower detectors, an electromagnetic
calorimeter and a hadronic calorimeter. Muons are identified by a
system composed of alternating layers of iron and multiwire
proportional chambers.

The online event selection is performed by a trigger~\cite{LHCb-DP-2012-004}, 
which consists of a hardware stage, based on information from the calorimeter and muon
systems, followed by a software stage, which applies a full event
reconstruction.
At the hardware trigger stage, events are required to have a muon with high \pt or a
hadron, photon or electron with high transverse energy in the calorimeters. 
 
A new scheme for the LHCb software trigger was introduced for LHC Run 2. 
Alignment and calibration are performed in near 
real-time~\cite{Dujany:2015lxd}
and updated constants 
are made available for the trigger. The same
alignment and calibration information is propagated to the offline reconstruction, 
ensuring high-quality particle identification (PID) and
consistent information between the 
trigger
and offline software. The larger timing budget available in the trigger compared to 
that available in
Run 1 also results in the convergence of the online and offline track reconstruction, 
such
that offline performance is achieved in the trigger. The identical performance of the 
online
and offline reconstruction offers the opportunity to perform physics analyses 
directly using
candidates reconstructed in the trigger~\cite{Benson:2015yzo}.  
 
In the simulation, $pp$ collisions are generated using
\pythia~\cite{Sjostrand:2006za,*Sjostrand:2007gs} 
with a specific \lhcb
configuration~\cite{LHCb-PROC-2010-056}.  Decays of hadronic particles
are described by \evtgen~\cite{Lange:2001uf}, in which final-state
radiation is generated using \photos~\cite{Golonka:2005pn}. The
interaction of the generated particles with the detector, and its response,
are implemented using the \geant
toolkit~\cite{Allison:2006ve, *Agostinelli:2002hh} as described in
Ref.~\cite{LHCb-PROC-2011-006}.

\section{Data samples and outline of analysis method}
\label{sec:Outline}
In the analysis, the decays \dppp and \dsppp are used to look for the presence of $\eta$ and $\eta^\prime$ resonances
in the \pp mass spectra, which could come from the known decays
$D_{(s)}^+ \to \pi^+\etas$ (inclusion of charge-conjugate 
modes is implied throughout). The data samples comprise about 
$25$ million each of \dppp and \dsppp decays, from integrated 
luminosities of 3.0\,fb$^{-1}$ 
of $pp$ collision data 
recorded by LHCb in LHC Run 1 and 0.3\,fb$^{-1}$ recorded in 2015 during Run~2.  

For $N(\etas)$ observed $\eta^{(\prime)}$ signal decays in
the $\pi^+\pi^-$ mass spectrum from a total of $N(D_{(s)}^+)$ mesons
reconstructed in the $\pi^+\pi^+\pi^-$ final state, the measured
branching fraction would be
\begin{equation}
\BRof\etaspp = \frac{N(\etas)}{N(D_{(s)}^+)}\times\frac{{\cal B}(\dpsppp)}{{\cal B}(D_{(s)}^+ \to \pi^+\etas)}\times\frac{1}{\epsilon(\eta^{(\prime)})}\,,
\label{eq:BF}
\end{equation}
where $\epsilon(\etas)$ accounts for any variation of efficiency 
with \pp 
mass, as discussed in Sec.~\ref{sec:efficiency}. The values of $N(D_{(s)}^+)$ and $N(\etas)$ and their uncertainties are obtained from 
fits to the $\pi^+\pi^-\pi^+$ and $\pi^+\pi^-$ mass spectra
of the selected \dpsppp candidates;
the branching fractions
${\cal B}(\dpsppp)$ and ${\cal B}(D_{(s)}^+ \to \pi^+\etas)$ and their
uncertainties are taken from 
Ref.~\cite{PDG2014}; and the relative efficiency factors, $\epsilon$, are
obtained from simulations.
Since the analysis starts from a given number of
selected \dpsppp 
decays, there are no normalisation channels.
All selections are finalised and 
expected
sensitivities are evaluated before the $\eta$ and $\eta^\prime$ 
signal 
regions in the $\pi^+\pi^-$ mass spectra are examined.

\section{Event selection}
\label{selection}

The event selection comprises an initial stage in which relatively loose criteria
are applied to select samples of candidate \dpsppp decays. A boosted decision tree (BDT)~\cite{Roe}
is then used to further suppress backgrounds. 

Candidate \dpsppp decays are required
to have three good quality tracks,
each with \pt greater than $250$\,\mevc, consistent with coming from a vertex that is 
displaced from any PV in the event. Loose particle identification
criteria are applied, requiring the tracks 
to be consistent with the pion hypothesis. The three-track system
is required to have total charge $\pm e$, its invariant mass must be in the range 
$1820$--$2020$\,\mevcc, and its combined momentum vector 
must be consistent with 
the direction from a PV to the decay vertex. The invariant
mass of opposite-sign candidate pion pairs is required to be in the range $300$--$1650\mevcc$; this removes backgrounds where a random pion is associated with a vertex from either a 
$\gamma \to e^+e^-$ 
conversion, in which both electrons are misidentified as pions, or from a $D^0 \to K^-\pi^+$ decay, where the kaon is misidentified as
a pion.

The BDT has six input variables for each of the 
tracks, 
together with three variables related to the quality of
the decay vertex and the association of the $D^+_{(s)}$ candidate with the PV. The track 
variables are related to track fit quality, particle identification probabilities and the quality of the
track association
to the decay vertex. The BDT is trained using 
a sample of $820\,000$ simulated \dppp
events for the signal, generated uniformly in phase space,
and about $10^7$ background candidates obtained from
sidebands of width $20 \mevcc$ on each side of the \dppp mass peak
in the data. 

\begin{figure}[tb]
  \begin{center}
    \includegraphics[width=0.8\linewidth]{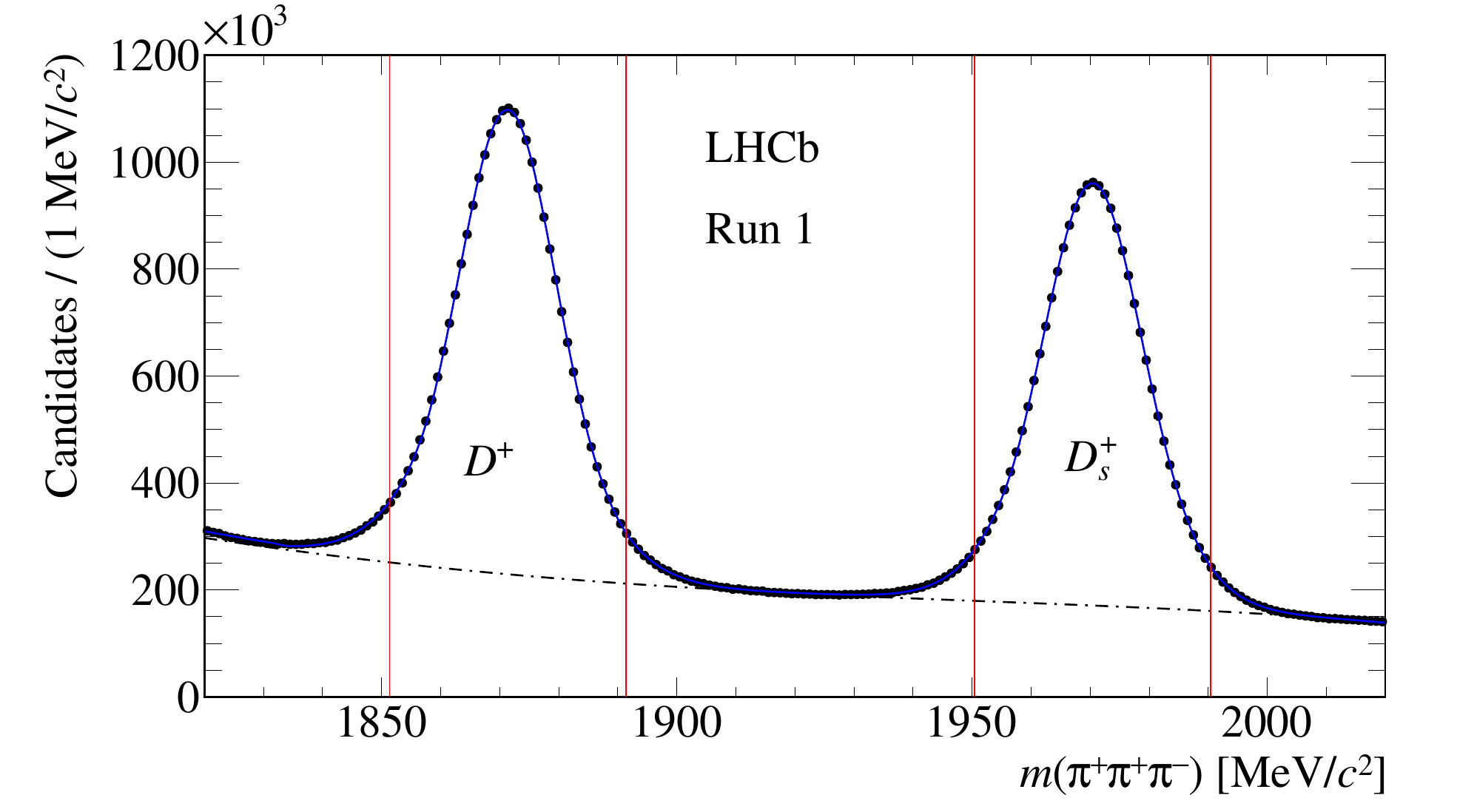}
      \includegraphics[width=0.8\linewidth]{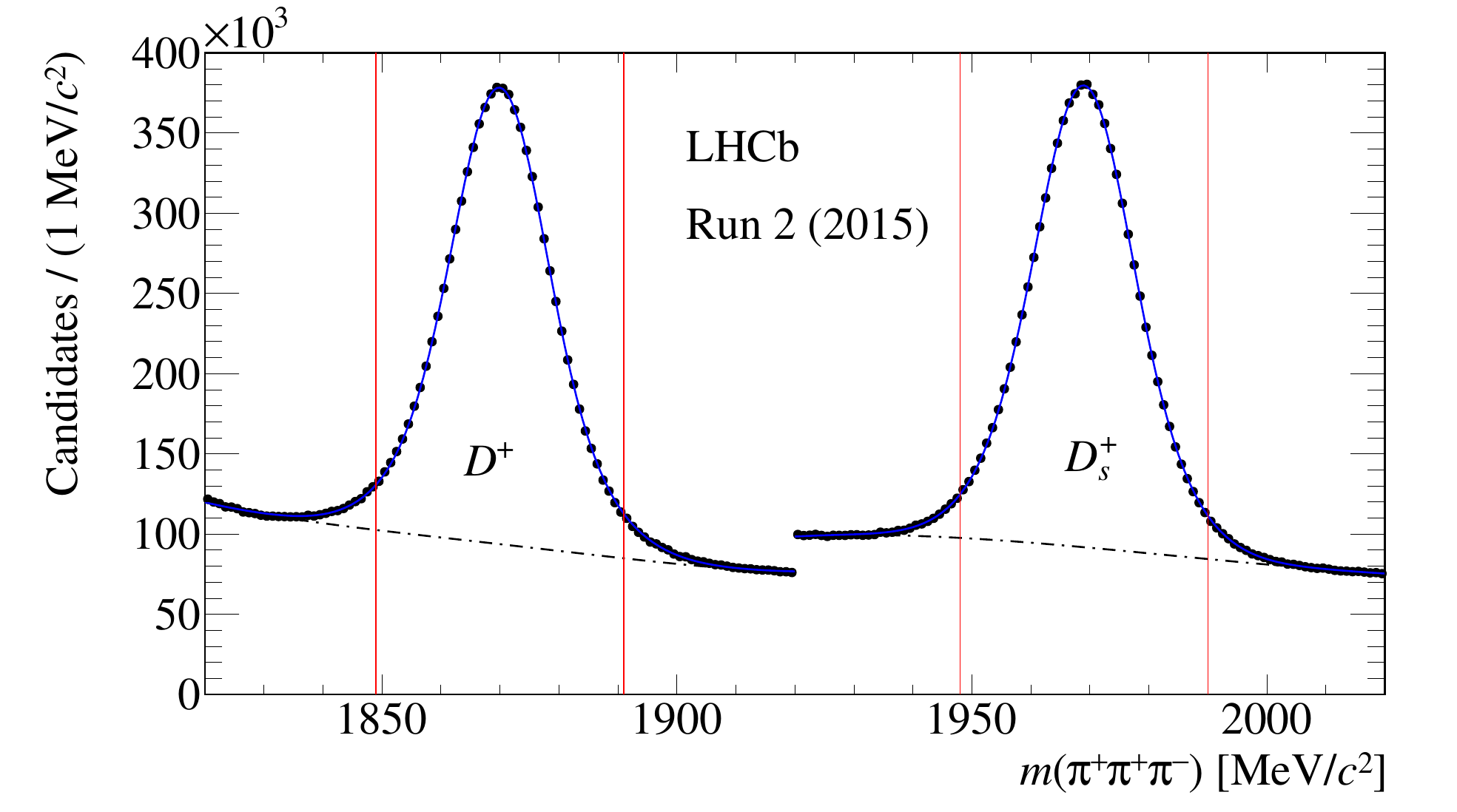}
  \end{center}
  \caption{
    \small 
    Mass spectra of selected \protect\dpsppp candidates, after the BDT selections, for (top) Run 1 and
    (bottom) Run 2 data, with the results from the fits superimposed. The dot-dashed lines show the
    total fitted backgrounds, and the vertical lines indicate 
    the optimised $D^+_{(s)}$ signal regions. The discontinuity in the Run 2 spectrum comes
from the fact that the trigger has two separate
output streams and
there are different BDT selections for \protect\Dp and \protect\Dsp.}
  \label{fig:Mass3pi}
\end{figure}

The selection criteria for the BDT output value and \threepi signal mass 
windows are 
simultaneously optimised to
maximise the statistical significance of the $D_{(s)}^+$ signals, $N_{\mathrm {sig}}/\sqrt {{N_{\mathrm {sig}}}+N_{\mathrm {bkg}}}$,
where $N_{\mathrm {sig}}$ is the number of $D_{(s)}^+$ signal decays and $N_{\mathrm {bkg}}$ is the number
of background events within the signal mass windows. The BDT 
selection gives signal efficiencies of $90\%$ while rejecting
about $60\%$ of the backgrounds. The optimum mass 
selection ranges are $\pm20\mevcc$ for both the \Dp and \Dsp 
peaks
in Run 1 and $\pm21\mevcc$ for both peaks in Run 2. 

Figure~\ref{fig:Mass3pi} shows the $\threepi$ mass spectra for Runs 1 and 2, after 
the BDT selection. The discontinuity in the Run 2 spectrum comes
from the fact that the trigger has two separate
output streams and
there are different BDT cuts for \Dp and \Dsp. 
The yield per fb$^{-1}$ is larger in Run 2  than in Run 1 by a factor $3.3$, arising from the larger  
cross-section~\cite{LHCb-PAPER-2015-041}, and from a higher trigger efficiency for charm.
The curves in Fig.~\ref{fig:Mass3pi} show the results of fits to the spectra in which 
each peak is parametrised by the sum of a double-sided Crystal Ball function~\cite{Skwarnicki:1986xj} and
a Gaussian function, while a fourth-order polynomial is used for the combinatorial background. All 
shape and yield parameters are allowed to vary in the fits. The fits
also include components for contributions from $D_s^+ \to K^+\pi^+\pi^-$ 
decays, where the kaon
is misidentified as a pion, and from $D_s^+ \to \pi^+\pi^+\pi^-\pi^0$ and 
$D^+_{(s)} \to \pi^+\eta^{(\prime)}$ with $\eta^{(\prime)} \to \pi^+\pi^-\gamma$. The
yields for these last components, the shapes for which are obtained from simulation, are found to be small. The total \dpsppp signal yields in the optimised mass windows, 
summed over Run 1 and Run 2 data, are $2.49 \times 10^7$ for \Dp 
and $2.37 \times 10^7$ for \Dsp, with backgrounds of $1.38 \times 10^7$ and $1.08 \times 10^7$, respectively, within the same mass windows.
Uncertainties of
$\pm2 \%$, corresponding to the maximum values of the fit residuals, are assigned to each total yield to account for imperfections in 
the fits to the mass spectra.
To improve the \pp mass resolution, a kinematic fit~\cite{Hulsbergen:2005pu} is 
performed on the selected $D^+_{(s)}$ candidates, with the three
tracks constrained to a common vertex, the $\pi^+\pi^+\pi^-$ mass constrained to the known $D^+_{(s)}$ mass, and the $D^+_{(s)}$ candidate constrained to come from the PV.

\section{Limits on the {\boldmath {\protect\etaspp}} branching fractions}

\subsection{Mass spectra for {\boldmath {\protect\pippim}}}

\begin{figure}[!t]
\centering
\includegraphics[width=0.9\linewidth]{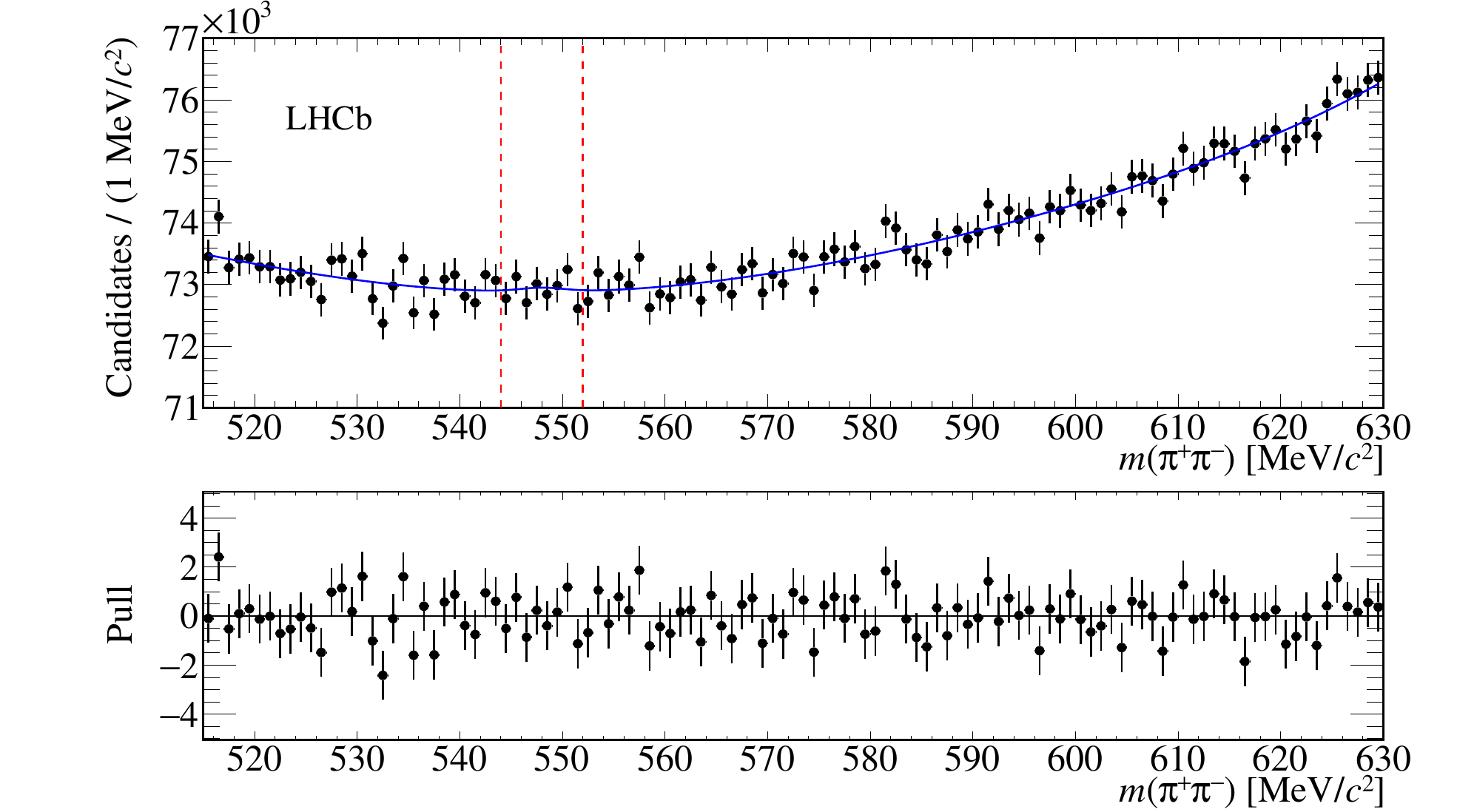}
\caption{\small The $\pi^+\pi^-$ invariant mass distribution in the $\eta$ mass fitting region 
from the sum of the four samples, showing also the sum of the fitted curves and the pulls. The vertical dashed lines indicate the $\eta$ signal region.}
\label{fig:fiteta_total}
\end{figure}

\begin{figure}[!t]
\centering
\includegraphics[width=0.9\linewidth]{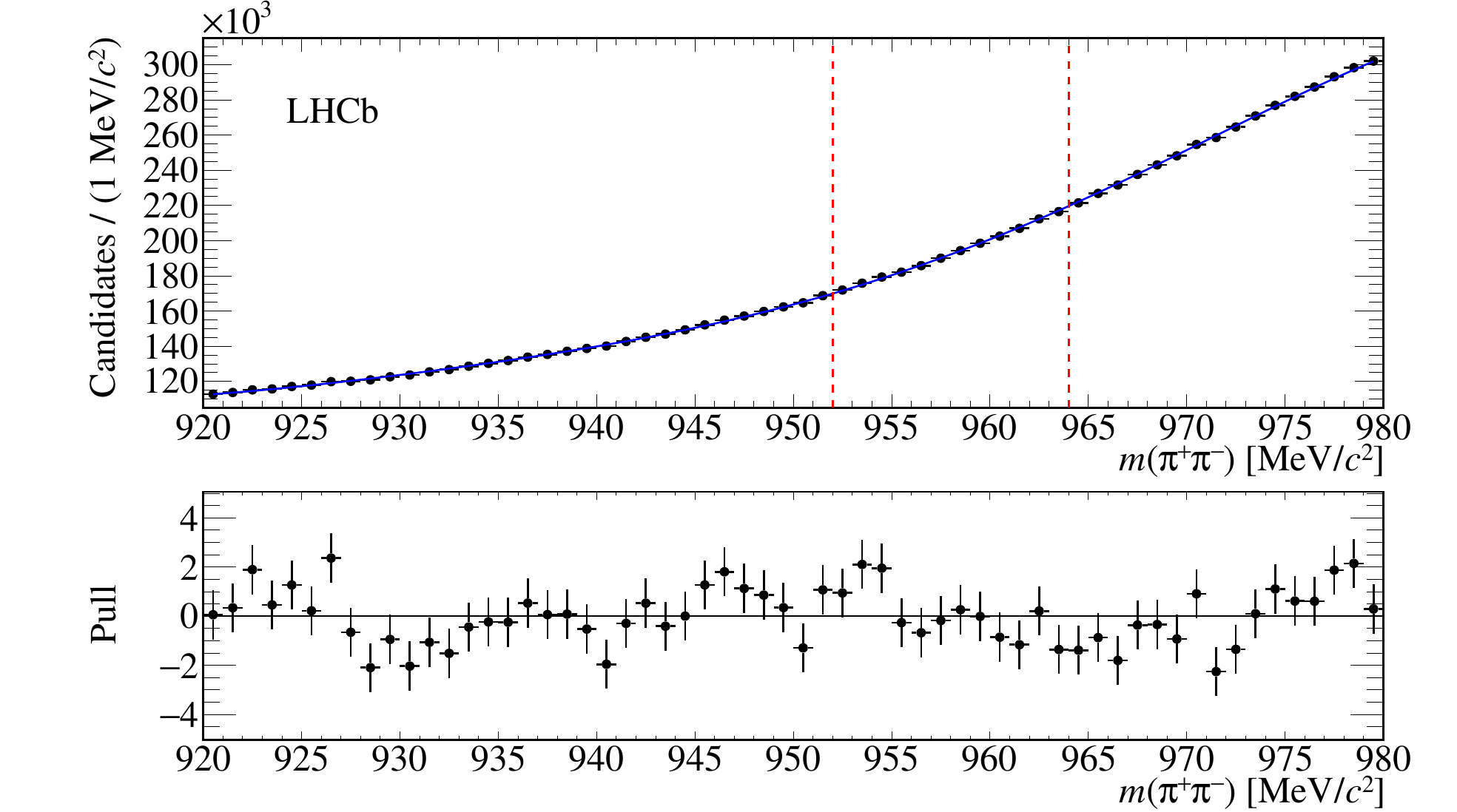}
\caption{\small The $\pi^+\pi^-$ invariant mass distribution in the $\eta^\prime$ mass fitting region 
from the sum of the four samples, showing also the sum of the fitted curves and the pulls. The vertical dashed lines indicate the $\eta^\prime$ signal region.}
\label{fig:fitetap_total}
\end{figure}

For each of the $\eta$ and $\eta^\prime$ resonances there 
are four separate \pp mass
spectra, from the $D^+$ and the $D^+_s$ for each
of Runs 1 and 2.
Figures~\ref{fig:fiteta_total} and~\ref{fig:fitetap_total} show the 
sums of the four \pp mass spectra for the $\eta$ and $\eta^\prime$ mass
fitting ranges, which are chosen to avoid the peaks from the \KS,
$\rho(770)^0$ and $f_0(980)$ mesons. 
The fitting ranges are
515--630\,\mevcc for the $\eta$ and 920--964\,\mevcc 
for the $\eta^\prime$.
The vertical dashed lines indicate
the signal regions, which cover the intervals 
544--552\mevcc for 
the $\eta$ and
952--964\mevcc for the $\eta^\prime$, in each case approximately 
$\pm 2$ times the \pp mass resolution. Simulation studies of the decays
$\eta^{(\prime)} \to \pi^+\pi^-\gamma$, using the matrix element given in 
Ref.~\cite{Adlarson:2011xb},
show that the contributions from these channels are small and do not peak in the fitting 
ranges. They are therefore considered as part of the background, which is 
parametrised by a polynomial function (see Sect.~\ref{sec:fits}).

Expected signal \pp mass line shapes for \epp and \eppp are obtained
from simulations. In both cases a double Gaussian shape is found to
describe the signal well, with mass resolutions of 
$2.3$\mevcc for the
$\eta$ mass region and $3.2$\mevcc for the $\eta^\prime$ region. These results
are calibrated by comparing the $\eta$ mass resolution from the simulation with that for reconstructed $\KS \to \pp$ decays
from background $D^+_{(s)} \to \KS \pi^+$ events in the
data, before the kinematic fits to the 
$D^+_{(s)}$ candidates. The differences, which are $5\%$ in Run 1 and $10\%$ in Run 2,
are taken as the systematic uncertainties on the \pp mass resolution for both the 
$\eta$ and $\eta^\prime$ mass
ranges.

\subsection{Relative efficiency as a function of {\boldmath \protect\pippim} mass}\label{sec:efficiency}

The relative efficiency factors in Eq.~\ref{eq:BF} are obtained from simulation. 
Fully simulated \pp mass spectra from $D^+ \to \pi^+\pi^+\pi^-$ decays 
for Run 1
are divided by the generated spectra to give the
relative efficiency as a function of the \pp mass. The efficiency is 
highest at large $\pi^+\pi^-$ masses, 
mainly due to the effects of the hardware and software triggers. The relative 
efficiencies in Run 1 data are found to be
$\epsilon(\eta) = 0.85 \pm 0.01$ and 
$\epsilon(\eta^\prime) = 1.01 \pm 0.01$, where the uncertainties come from the simulation sample size. The relative efficiencies for Run 2 are found to be statistically
compatible with those for Run 1, through a comparison of 
the \pp mass spectra from the $D^+$ and $D_s^+$ signal candidates in the data. 
An additional systematic uncertainty of $2\%$ is assigned to the Run 2 relative 
efficiencies, corresponding to the maximum 
difference between the mass spectra.

\subsection{Sensitivity studies}
\label{sec:fits}

\begin{figure}[!t]
\centering
\includegraphics[width=0.8\textwidth]{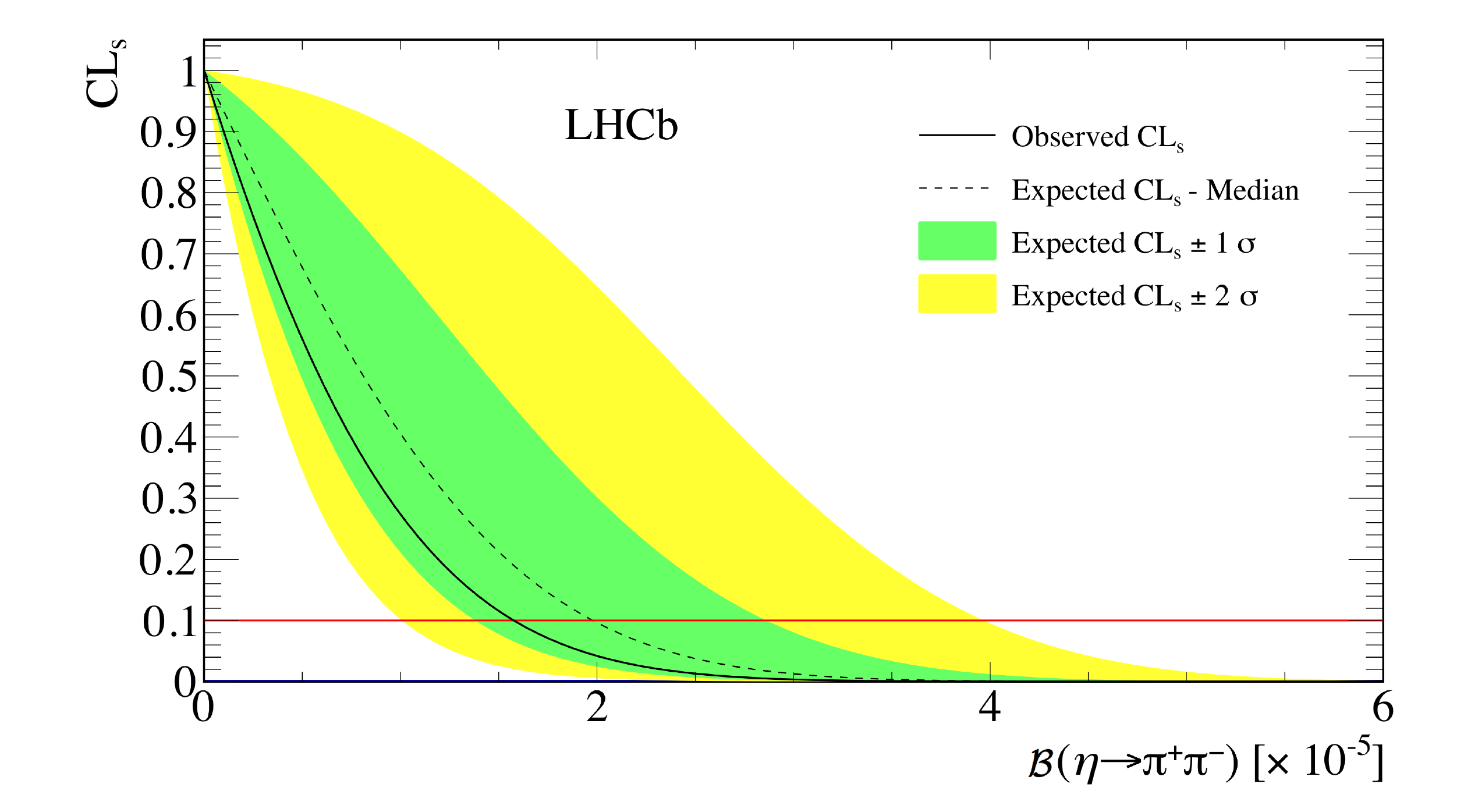}
\caption{\small Values of \protect\CLs as a function of \protect\BRof{\epp}.
The expected variation is shown by the dashed line, with the $\pm 1\sigma$ and $\pm 2\sigma$ regions shaded. The observed variation is shown by the 
solid line, while the horizontal line indicates the $90\%$ confidence level.}
\label{fig:CLs_eta}
\end{figure}

\begin{figure}[!t]
\centering
\includegraphics[width=0.8\textwidth]{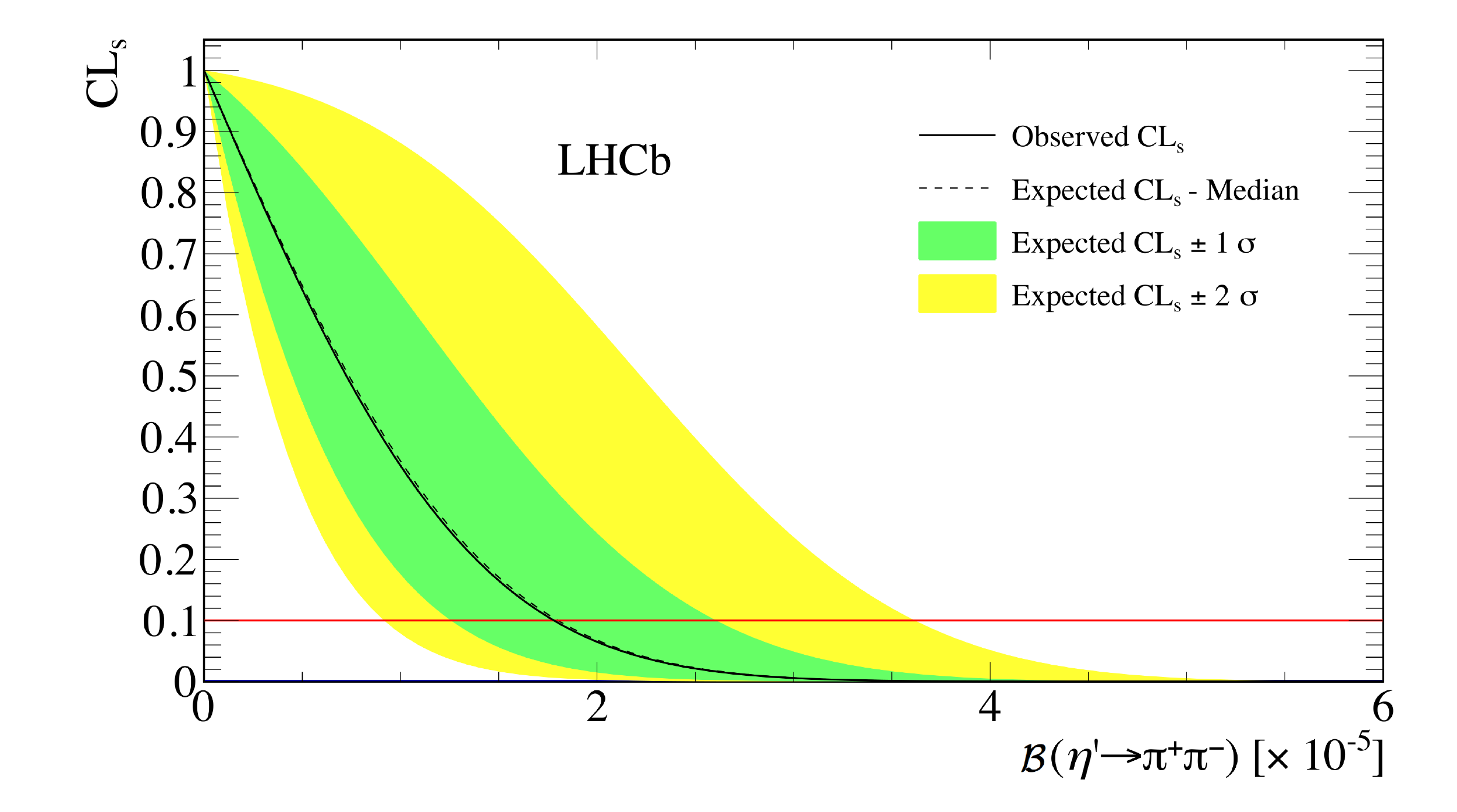}
\caption{\small Values of \protect\CLs as a function of \protect\BRof{\eppp}.
The expected variation is shown by the dashed line, with the $\pm 1\sigma$ and $\pm 2\sigma$ regions shaded. The observed variation is shown by the 
solid line, which almost overlays the dashed line, while the horizontal line indicates the $90\%$ confidence level.}
\label{fig:CLs_etaprime}
\end{figure}

In order to measure the sensitivity of the analysis, each \pp mass spectrum is fitted with a fourth-order polynomial, initially with the signal 
regions excluded. The signal regions are then populated with pseudo data, 
generated according to the fitted polynomial functions, with Gaussian fluctuations. 
Each spectrum is then fitted again with the sum of a fourth-order polynomial plus 
the $\eta^{(\prime)}$ signal function, and Eq.~\ref{eq:BF} is then used to obtain branching
fractions measured with the pseudo data. As expected, these branching fractions
are consistent with zero.
Expected upper limits on the branching fractions are
obtained using the \CLs method~\cite{CLsMethod}.
In each case, \CLs values are obtained using
the products of the likelihood functions for the four individual spectra.
Systematic uncertainties are included, but have no effect on the results,
which are shown in Fig.~\ref{fig:CLs_eta}
for the $\eta$ and in Fig.~\ref{fig:CLs_etaprime} for the $\eta^\prime$.
Expected limits at $90\%$ CL are $\BRof{\epp}<2.0 \times 10^{-5}$
and $\BRof{\eppp}<1.8 \times 10^{-5}$.
 
\subsection{Observed limits on the branching fractions}
 
The procedures outlined above are then applied to the observed mass
spectra, \ie with the pseudo data in the signal ranges replaced by
the observed data. The sums of the fits to the four spectra for the
$\eta$ and $\eta^\prime$ are shown as the solid curves in 
Figs.~\ref{fig:fiteta_total} and~\ref{fig:fitetap_total}.
The results are consistent among the four mass spectra for each meson. 
Weighted average branching fractions are measured to be
$\BRof{\epp} = (-1.1 \pm 1.8) \times 10^{-5}$
and $\BRof{\eppp} = (0.8 \pm 1.6) \times 10^{-5}$. Although the simple,
unweighted sum of the fits to the \pp mass spectra in Fig.~\ref{fig:fiteta_total} shows a small, but insignificant, positive
yield for the $\eta$, the weighted average branching fraction 
\BRof{\epp} is dominated 
by a negative value in the Run 1 \Ds sample.
 
Since there is no evidence for any signal, the \CLs method is used, as for the pseudo data, to obtain observed upper limits on the branching fractions. 
Figures~\ref{fig:CLs_eta} and~\ref{fig:CLs_etaprime} show the observed \CLs values as functions of the branching fractions. Limits obtained are 
$$\BRof{\epp}< 1.6 \times 10^{-5},$$
$$\BRof{\eppp}<1.8 \times 10^{-5},$$ 
both at $90\%$ CL, in good agreement with the expected limits.

\section{Conclusions}

A new method is introduced to search for the decays \epp and 
$\eta^\prime(958) \to \pp$,
which would violate \CP symmetry in the strong interaction. The method
relies on the copious production of charm mesons at \lhcb, and will
improve in sensitivity as more data are collected at the LHC. With the
LHC Run 1 data and data from the first year of Run 2, the limit obtained on
the branching fraction for the decay \epp is comparable to the existing limit, while
that for \eppp is a factor three better than the previous limit.
	
\section*{Acknowledgements}

\noindent We express our gratitude to our colleagues in the CERN
accelerator departments for the excellent performance of the LHC. We
thank the technical and administrative staff at the LHCb
institutes. We acknowledge support from CERN and from the national
agencies: CAPES, CNPq, FAPERJ and FINEP (Brazil); NSFC (China);
CNRS/IN2P3 (France); BMBF, DFG and MPG (Germany); INFN (Italy); 
FOM and NWO (The Netherlands); MNiSW and NCN (Poland); MEN/IFA (Romania); 
MinES and FASO (Russia); MinECo (Spain); SNSF and SER (Switzerland); 
NASU (Ukraine); STFC (United Kingdom); NSF (USA).
We acknowledge the computing resources that are provided by CERN, IN2P3 (France), KIT and DESY (Germany), INFN (Italy), SURF (The Netherlands), PIC (Spain), GridPP (United Kingdom), RRCKI and Yandex LLC (Russia), CSCS (Switzerland), IFIN-HH (Romania), CBPF (Brazil), PL-GRID (Poland) and OSC (USA). We are indebted to the communities behind the multiple open 
source software packages on which we depend.
Individual groups or members have received support from AvH Foundation (Germany),
EPLANET, Marie Sk\l{}odowska-Curie Actions and ERC (European Union), 
Conseil G\'{e}n\'{e}ral de Haute-Savoie, Labex ENIGMASS and OCEVU, 
R\'{e}gion Auvergne (France), RFBR and Yandex LLC (Russia), GVA, XuntaGal and GENCAT (Spain), Herchel Smith Fund, The Royal Society, Royal Commission for the Exhibition of 1851 and the Leverhulme Trust (United Kingdom).

\addcontentsline{toc}{section}{References}
\setboolean{inbibliography}{true}
\bibliographystyle{LHCb}
\bibliography{main,LHCb-PAPER,LHCb-CONF,LHCb-DP,LHCb-TDR,ourbib}

\newpage

 
\newpage
\centerline{\large\bf LHCb collaboration}
\begin{flushleft}
\small
R.~Aaij$^{40}$,
B.~Adeva$^{39}$,
M.~Adinolfi$^{48}$,
Z.~Ajaltouni$^{5}$,
S.~Akar$^{6}$,
J.~Albrecht$^{10}$,
F.~Alessio$^{40}$,
M.~Alexander$^{53}$,
S.~Ali$^{43}$,
G.~Alkhazov$^{31}$
% ...

{\footnotesize \it
% ...
$ ^{5}$Clermont Universit{\'e}, Universit{\'e} Blaise Pascal, CNRS/IN2P3, LPC, Clermont-Ferrand, France\\
$ ^{6}$CPPM, Aix-Marseille Universit{\'e}, CNRS/IN2P3, Marseille, France\\
% ...
$ ^{10}$Fakult{\"a}t Physik, Technische Universit{\"a}t Dortmund, Dortmund, Germany\\
% ...
$ ^{31}$Petersburg Nuclear Physics Institute (PNPI), Gatchina, Russia\\
% ...
$ ^{39}$Universidad de Santiago de Compostela, Santiago de Compostela, Spain\\
$ ^{40}$European Organization for Nuclear Research (CERN), Geneva, Switzerland\\
% ...
$ ^{43}$Nikhef National Institute for Subatomic Physics, Amsterdam, The Netherlands\\
% ...
$ ^{48}$H.H. Wills Physics Laboratory, University of Bristol, Bristol, United Kingdom\\
% ...
$ ^{53}$School of Physics and Astronomy, University of Glasgow, Glasgow, United Kingdom\\
% ...
}
\end{flushleft} 
\end{document}